\documentclass[a4paper,UKenglish,cleveref, autoref, thm-restate]{lipics-v2021}
\usepackage{caption}
\usepackage{cleveref}

\hideLIPIcs  

\title{UAIC\_Twin\_Width: An Exact yet Efficient Twin-Width Algorithm} 

\author{Andrei Arhire}{Alexandru Ioan Cuza University of Iași, Romania\and }{andrei.arhire@info.uaic.ro}{}{}

\author{Matei Chiriac}{Saarland University, Germany\and }{chiriac.matei@gmail.com}{}{}

\author{Radu Timofte}{Computer Vision Lab, CAIDAS \& IFI, \and University of Würzburg, Germany \and Alexandru Ioan Cuza University of Iași, Romania}{radu.timofte@uni-wuerzburg.de}{}{}

\Copyright{Andrei Arhire} 



\category{} 

\authorrunning{A. Arhire, M. Chiriac, R. Timofte} 

\supplement{The source code is available on Zenodo \href{https://zenodo.org/record/8010483}{(10.5281/zenodo.8010483)} and GitHub \href{https://github.com/AndreiiArhire/PACE2023}{ (https://github.com/AndreiiArhire/PACE2023)}.}

\nolinenumbers 

\EventEditors{John Q. Open and Joan R. Access}
\EventNoEds{2}
\EventLongTitle{42nd Conference on Very Important Topics (CVIT 2016)}
\EventShortTitle{CVIT 2016}
\EventAcronym{CVIT}
\EventYear{2016}
\EventDate{December 24--27, 2016}
\EventLocation{Little Whinging, United Kingdom}
\EventLogo{}
\SeriesVolume{42}
\ArticleNo{23}

\begin{document}

\maketitle

\begin{abstract}

Twin-width is a recently formulated graph and matrix invariant that intuitively quantifies how far a graph is from having the structural simplicity of a co-graph. Since its introduction in 2020, twin-width has received increasing attention and has driven research leading to notable advances in algorithmic fields, including graph theory and combinatorics. The 2023 edition of the Parameterized Algorithms and Computational Experiments (PACE) Challenge aimed to fulfill the need for a diverse and consistent public benchmark encompassing various graph structures, while also collecting state-of-the-art heuristic and exact approaches to the problem. In this paper, we propose two algorithms for efficiently computing the twin-width of graphs with arbitrary structures, comprising one exact and one heuristic approach. The proposed solutions performed strongly in the competition, with the exact algorithm achieving the best student result and ranking fourth overall. We release our source code publicly to enable practical applications of our work and support further research.

\end{abstract}

\section{Introduction}

The concept of twin-width was introduced in \cite{bonnet2021twin} as a new measure of structural complexity for graphs, defined through sequences of vertex contractions that control the irregularities in adjacency that emerge during contraction. The authors initially proposed a fixed-parameter tractable (FPT) algorithm for first-order (FO) model checking on graphs of bounded twin-width, and subsequently developed a series of theoretical results expanding the structural and algorithmic understanding of the parameter (\cite{bonnet2022twin}, \cite{2b422d3fcd8e4be3953f891527adce3c}, \cite{bonnet2024twin}, \cite{bonnet2022twin4}, \cite{bonnet2024twin2}, \cite{bonnet2022twin2}, \cite{bonnet2022twin3}). The idea of twin-width has rapidly attracted significant attention in theoretical computer science, inspiring numerous studies on its algorithmic (\cite{bonnet2022twin10}, \cite{berge2021deciding}), combinatorial (\cite{dreier2022twin}, \cite{pilipczuk2023graphs}), and logical~(\cite{bonnet2024twin20}, \cite{gajarsky2022stable}) aspects. Although deciding whether a graph has twin-width at most four is NP-complete~\cite{berge2021deciding}, a wide range of graph classes are known to have bounded twin-width, including trees and graphs of bounded tree-width, bounded rank-width graphs, subgraphs of $d$-dimensional grids, map graphs, and unit interval graphs. Among various graph families, the class of planar graphs has seen substantial theoretical progress and active research over recent years, with the known upper bound on their twin-width gradually improving (\cite{hlinveny2022twina}, \cite{jacob2022bounding}, \cite{bekos2022graph}) from 583, as established in \cite{bonnet2022reduced}, to 8 in \cite{hliněný2023twinwidthplanargraphs8}, while a lower bound of 7 was reported~in~\cite{kral2022planar}.

As a result of these theoretical advances, the establishment of a public benchmark encompassing graphs from diverse classes and sizes has become a pressing requirement.  The PACE Challenge \cite{pacechallengeWebsite} is an annual competition that aims to close the gap between theoretical research on parameterized algorithms and their practical applications since 2015. The 2023 edition \cite{bannach2023pace} focused on efficient twin-width computation, encouraging the development of state-of-the-art exact and heuristic approaches on a new, publicly available, and diverse benchmark introduced by the organizers. Our contribution consists of developing two solvers for computing the twin-width of graphs with arbitrary structures. Specifically, one is a heuristic method designed to quickly find a good upper bound using local search, while the other is an exact solver based on dynamic programming that integrates the heuristic to improve efficiency. Both solvers were submitted to the PACE 2023 Challenge, and in the final global leaderboard \cite{PACE2023Results}, the heuristic solver ranked sixth, while the exact solver ranked fourth, achieving the best student results. The remainder of the paper is organized as follows. Section 2 presents the preliminaries. Section 3 describes the algorithms implemented by the solvers. Section 4 discusses the upper and lower bounds used. Section 5 provides the technical details, while the conclusions are drawn in Section 6.

\section{Preliminaries}
\label{sec:typesetting-summary}

Consider a trigraph $G = (V, B, R)$ consisting of a set of vertices $V$, black edges $B$, and red edges $R$. Here, $B$ and $R$ are subsets of $\binom{V}{2}$, and they do not have any common elements ($B \cap R = \emptyset$). Without any specific indication, a graph $H = (V, E)$ is treated as the trigraph $H' = (V, E, \emptyset)$. In this context, we define the red degree of a vertex as the number of incident red edges connected to that vertex. When two vertices are compressed (refer to \cref{fig:vertex_contraction}), a new vertex is formed by considering the edges associated with the original vertices, and subsequently, the original vertices are removed from the graph.

\begin{center}
\begin{minipage}{0.75\textwidth}
    \includegraphics[width=\linewidth]{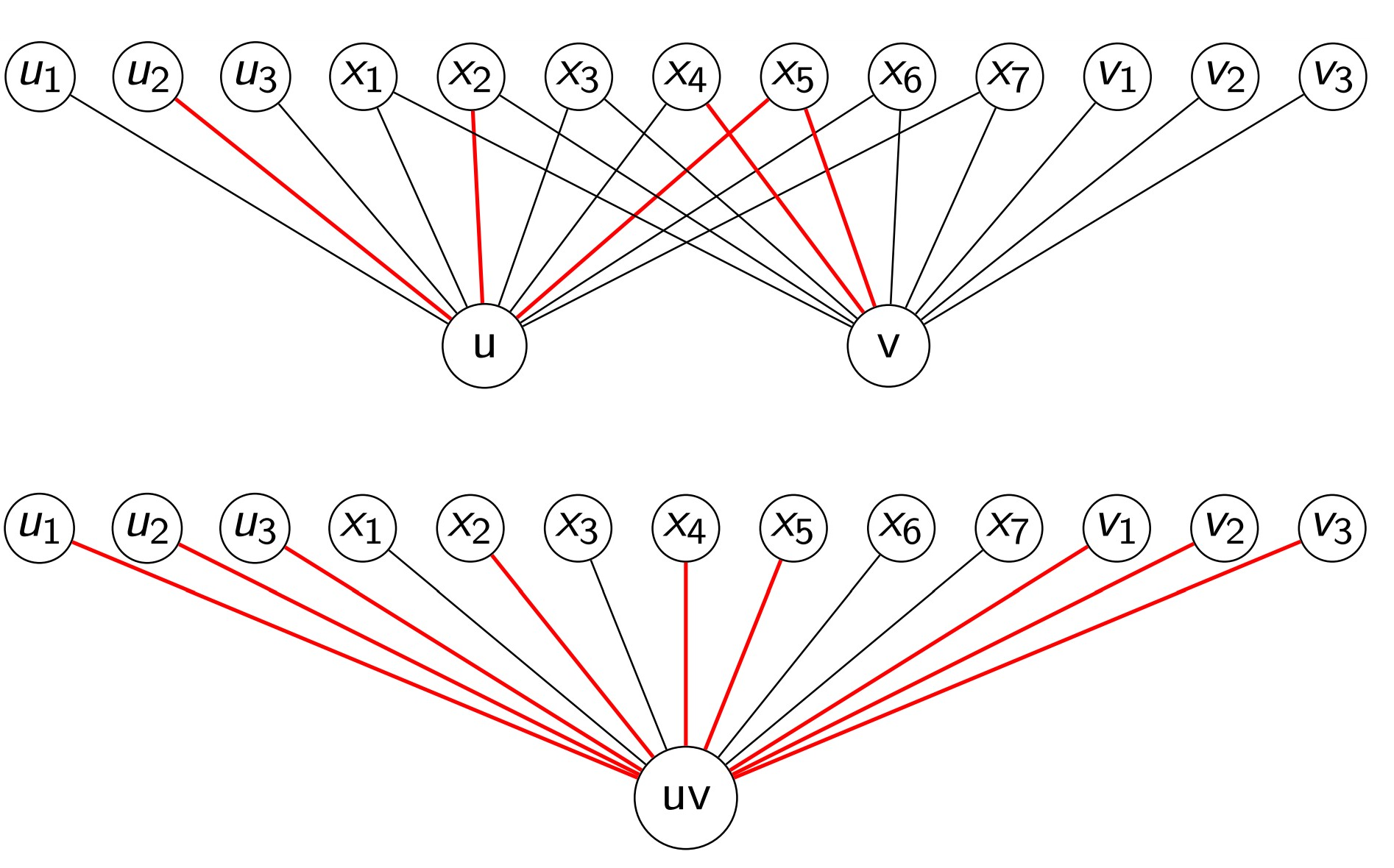}
    \captionof{figure}{Vertex contraction example showing how red and black edges are updated when merging vertices $u$ and $v$ into $uv$.}
    \label{fig:vertex_contraction}
\end{minipage}
\end{center}

Formally, the union of vertices $x$ and $y$, with $B_x$, $B_y$, $R_x$, and $R_y$ representing the sets of incident black and red edges, gives rise to a new vertex $z$ with the following definitions:

\begin{itemize}
\item A vertex $u \in R_z$ if $u \in { B_x \cup B_y \cup R_x \cup R_y } \setminus {B_x \cap B_y}$.
\item A vertex $u \in B_z$ if $u \in B_x \cap B_y$.
\end{itemize}

\begin{center}
\begin{minipage}{0.75\textwidth}
    \includegraphics[width=\linewidth]{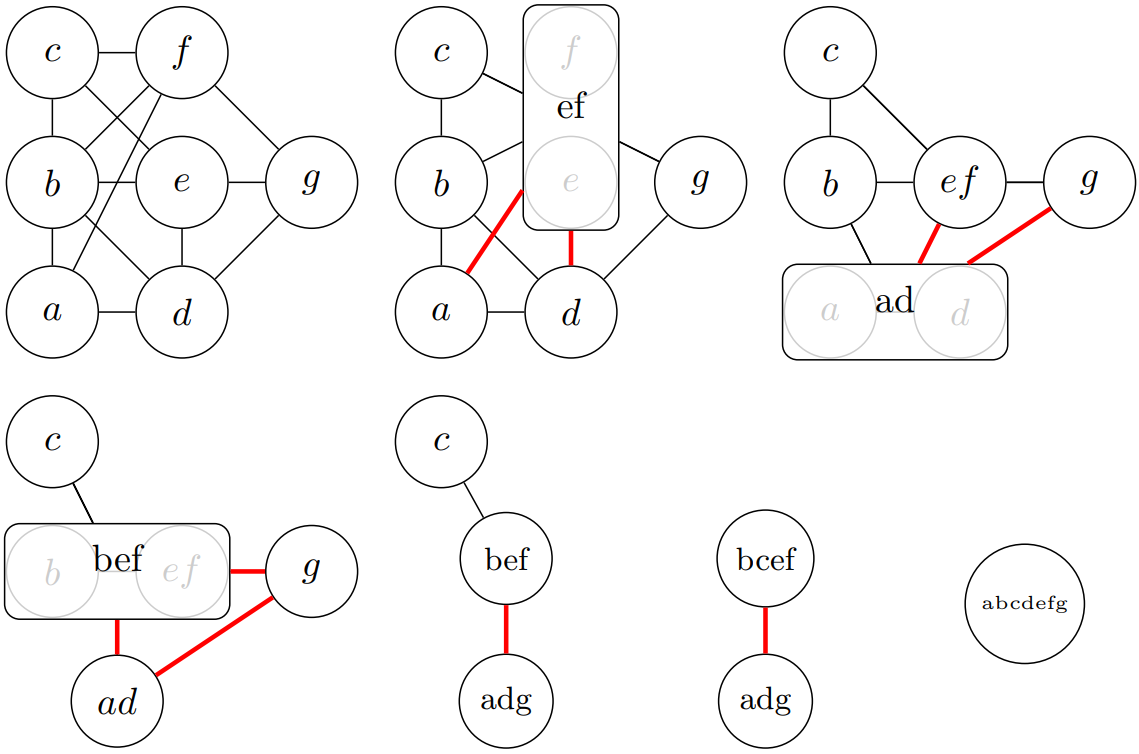}
    \captionof{figure}{Contraction sequence showing how the red and black edges are updated. The resulting graph has twin-width at most 2. Adapted from \cite{berge2021deciding}.}
    \label{fig:contraction_seq}
\end{minipage}
\end{center}

According to the requirements of the challenge format, the solver assigns the same label to $z$ as that of $x$. Following this process of reducing the graph to a single vertex through these operations, a sequence of $|V| - 1$ pairs of vertices is generated. The width of such a sequence is defined as the maximum red degree encountered at any vertex during the contraction process. The twin-width represents the minimum width achievable through a sequence of (refer to \cref{fig:contraction_seq}) that reduces the graph.

\section{Description of the Solver}

The solver can be divided into four stages:
\begin{enumerate}
\item Contract vertices that do not create any red edges when contracted;
\item Find an upper bound for the entire graph using a hill climbing algorithm;
\item Determine a lower bound for each connected component by using dynamic programming on multiple sub-graphs;
\item Compute the twin-width using dynamic programming. Evaluate components in descending order based on their number of vertices (expected twin-width). A component can yield a suboptimal solution as long as another component has a twin-width greater than or equal to its width.
\end{enumerate}
The core of the solver relies on a dynamic programming algorithm that iterates over subsets of potential solutions. A solution sequence can be visualized as a tree, where $x$ is the parent of $y$ if the pair $(x, y)$ is part of the sequence. Thus, the remaining vertex in the graph serves as the root. A partial solution can be seen as a set of trees.

A state in dynamic programming holds the following information:
\begin{itemize}
\item The sequence of contractions leading to the current solution;
\item The width of the current solution;
\item The set of trees determined by the sequence of contractions;
\item The initial state starts with an empty set, and the final state includes the twin-width value and the corresponding sequence. Transitioning between states is achieved by appending a pair of vertices to the partial solution.
\end{itemize}

To optimize the algorithm, the following observations are employed:
\begin{itemize}
\item All pairs used for state transitions consist of either adjacent vertices or vertices with at least one neighbor in common;
\item If there exists a pair of vertices whose contraction does not create any new red edges, it is optimal to contract that pair (as the vertices are twins);
\item To avoid ending up in non-optimal states multiple times, partial solutions are computed for $l$ pairs first, followed by partial solutions for $l + 1$ pairs, where $l < |V|$;
\item When comparing a state with $t$ trees ${ \alpha_1, \dots, \alpha_t }$ to a state with $t$ trees ${ \beta_1, \dots, \beta_t }$, where $\alpha_i$ and $\beta_i$ (for $1 \leq i \leq t$) have the same vertices but in different configurations, only the state with the lower width is retained. Thus, the removed state is guaranteed to be incapable of obtaining a better solution than the one kept;
\item Prior to running the algorithm, a low-width solution (upper bound) and a lower bound value are computed;
\item A state with a sequence width greater than or equal to the upper bound is excluded;
\item A state is excluded if the graph obtained by applying the sequence of contractions has a width lower or equal to the width of the state's sequence. For each state, a heuristic algorithm is used to find an upper-bound solution for the generated graph. If the upper bound is lower than or equal to the current width, appending the heuristic solution to the current sequence will result in a sequence with the same width;
\item A state with a width lower than the lower bound does not need to invoke the heuristic algorithm.
\end{itemize}
\section{Solver Bound Analysis}

An upper bound is calculated using two algorithms, namely the hill climbing algorithm ($H_a$) and the greedy algorithm ($G_a$), although they are used at different stages. On the entire graph, $H_a$ is utilized to determine an upper bound within a few minutes. Meanwhile, as part of the decision-making process, $G_a$ is used to determine whether a state should be excluded from the set of partial solutions.

$G_a$ constructs the sequence by gradually appending pairs of vertices to it, aiming to minimize the maximum red degree in the graph obtained after contracting the vertices.

$H_a$ begins with a primary solution generated by $G_a$ with a width of $w$. Let $d_i$, $1 \leq i \leq |V| - 1$, represent the maximum red degree of a vertex during the first $i$ contractions. Since $w$ represents the width, $d_{|V| - 1} = w$ at the very least. Consider $1 \leq p \leq |V| - 1$, where $d_{p} = d_{p+1} = \cdots = d_{|V| - 1}$ and $d_{p- 1} < w$.

$H_a$ explores multiple derived solutions and retains the one with the highest value of $p$, which indicates the solution that remains of width $w$ for as long as possible, preferably transitioning to width $w - 1$.

Let ${ (x_1, y_1), \cdots, (x_p, y_p), \cdots, (x_{|V|-1}, y_{|V|-1})}$ be a solution, where $p$ is as described above.

A potentially improved solution is obtained through the following steps:
\begin{itemize}
\item Randomly select integers $a$ and $b$ from the range $1$ to $p$, and randomly choose vertices $u$ and $v$ such that $u \neq y_i, 1 \leq i \leq a$ and $v \neq y_i, 1 \leq i \leq b$ (where $u$ is present in the graph after the $a$-th operation and $v$ is present after the $b$-th operation);
\item Update $x_a$ to be equal to $u$;
\item Swap $y_b$ with $v$, starting from line $p$;
\item Initiate $G_a$ from the partial solution consisting of the first $p$ (or fewer) pairs.
\end{itemize}
Once a specific number of derived solutions have been computed, the best-evaluated solution is designated as the primary solution. The complexity of $G_a$ is $\mathcal{O}(|V|^3 \cdot |E|)$, while $H_a$ has a complexity of $\mathcal{O}(|V|^3 \cdot |E| \cdot T)$, where $T$ represents the number of local searches performed.

In the case of public tests that are small enough for $G_a$ to execute rapidly, $H_a$ achieved the optimal answer in approximately one-fourth to one-third of the tests.

The twin-width of a graph is always greater than or equal to the twin-width of any of its sub-graphs. Exploiting this concept, we can determine a lower bound by utilizing the exact algorithm on various sub-graphs. When applied to public tests, this technique rapidly identifies the optimal solution for tests where the twin-width is at most three.

For small tests with a twin-width no greater than five, the upper bound often matches the lower bound. Consequently, it is unnecessary to execute the dynamic programming algorithm on the entire graph.

\section{Implementation Details}

The entire solution~\cite{arhire2023uaictwinwidth} was implemented in C++ for efficiency and does not employ parallelism, as this was prohibited by the committee. Several key announcements related to the challenge were made throughout the year and participants were able to submit their solutions over a period of three months (see \cref{fig:timeline}). In the early stage of the competition, the organizers provided a small set of 10 graphs to help participants familiarize themselves with the evaluation system. For this dataset, the runtime limit was set to 60 seconds. Each of~the~exact and heuristic tracks contains 200 instances, divided into 100 public instances for development and another 100 private instances from the same distribution, used for the final~evaluation. The participants could evaluate their solutions on the target platform~\cite{optilioPlatform} in both heuristic and exact tracks. A submitted solution in the exact track had to find the correct result within 30 minutes and a memory limit of 8 GB. The ranking was based on the number of instances solved, with the secondary criterion being the total time spent to obtain the results. For the heuristic track, the memory limit remained the same, but the time limit was reduced to 5 minutes, while the graphs were significantly larger. In this case,~the~final~score was computed for each instance based on the ratio between the obtained result and the best~known~result.   

\begin{center}
\begin{minipage}{0.95\textwidth}
    \includegraphics[width=\linewidth]{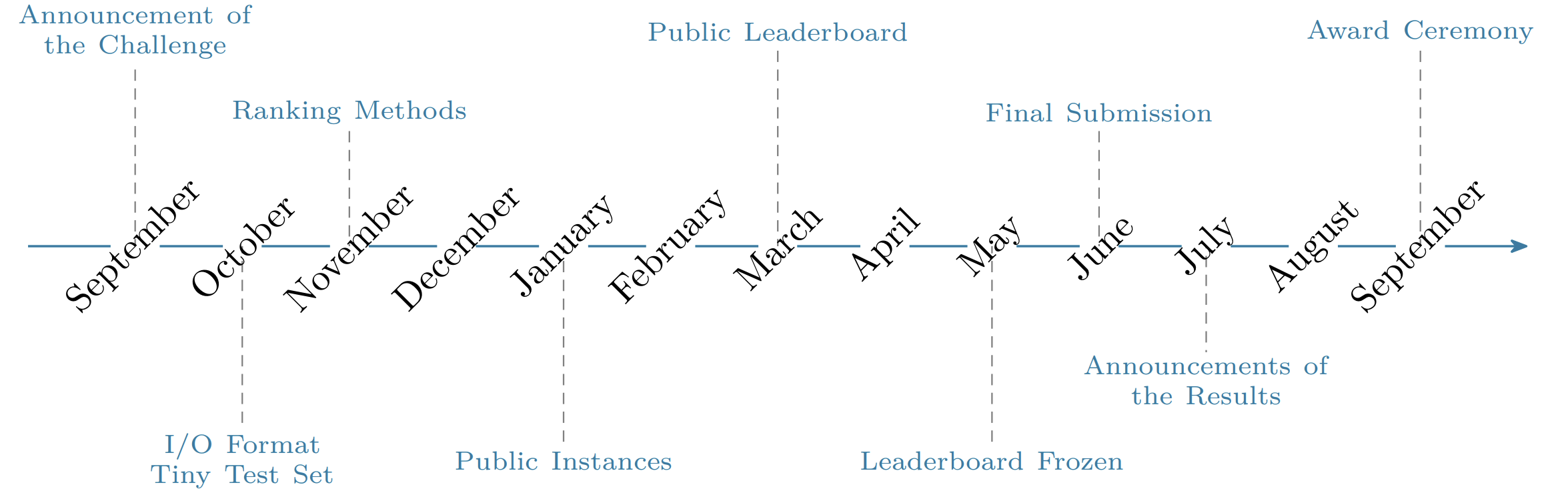}
    \captionof{figure}{Timeline of the 2023 edition of the PACE Challenge. Image adapted from \cite{bannach2023pace}.}
    \label{fig:timeline}
\end{minipage}
\end{center}

\section{Conclusion}

This paper presents an optimized exact algorithm for finding a contraction sequence of vertex pairs with optimal twin-width, based on dynamic programming and efficient heuristics. The strong performance \cite{PACE2023Results} demonstrated in the 2023 edition of the PACE Challenge shows that the proposed approach is both competitive and practical, and can be effectively used as a tool for analyzing various graph-structured data. Future work will focus on developing more accurate solutions for specific graph classes by leveraging the most recent theoretical~advances.



\bibliography{lipics-v2021-sample-article}

\appendix

\end{document}